\documentclass[]{aa}

\usepackage{graphicx,longtable,lscape}
\usepackage{txfonts,psfig}
\usepackage[]{natbib}
\def \ferg {erg cm$^{-2}$ s$^{-1}$}

\begin{document}

\title{The X-ray view of black-hole candidate \object{Swift J1842.5-1124} during its 2008 outburst}

\authorrunning{Zhao et al.}
\titlerunning {The X-ray view of \object{Swift J1842.5-1124}}

\author{H.-H. Zhao\inst{1}, S.-S. Weng\inst{1,2}\fnmsep\thanks{E-mail: wengss@njnu.edu.cn}, J.-L. Qu\inst{2}, J.-P.
Cai\inst{1}, Q.-R. Yuan\inst{1}} \institute{$^{1}$Department of Physics and
Institute of Theoretical Physics,
Nanjing Normal University, Nanjing 210023, China\\
$^{2}$Laboratory for Particle Astrophysics, Institute of High Energy Physics, Beijing 100049, China\\
}

\abstract {The spectral and temporal evolution during X-ray outbursts give
important clues on the accretion process and radiation mechanism in black-hole
X-ray binaries (BH XRBs).} {A set of \textit{Swift} and \textit{RXTE}
observations were executed to monitor the 2008 outburst of the black-hole
candidate \object{Swift J1842.5-1124}. We investigate these data to explore the
accretion physics in BH XRBs.} {We carry out a comprehensive spectral and
timing analysis on all the available pointing observations, including fitting
both X-ray spectra and power density spectra, measuring the optical and
near-ultraviolet flux density. We also search for correlations among the
spectral and timing parameters.} {The observed properties of \object{Swift
J1842.5-1124} are similar to other BH XRBs in many respects, for example the
hardness-intensity diagram and hardness-rms diagram. The type-C quasi-periodic
oscillations (QPOs) were observed as the source started to transit from the
low-hard state to the high-soft state. The frequency of QPOs correlate with
intensity and the hard component index, and anti-correlate with the hardness
and the total fractional rms. These relations are consistent with the
Lense-Thirring precession model. The estimated U-band flux changed with the
X-ray flux, while the flux density at the V band remained 0.26 mJy. These
results imply that the X-ray reprocessing or the tail of thermal emission from
the outer disk contributes a significant fraction of the U-band radiation;
alternatively, the companion star or the jet dominates the flux at longer
wavelengths.} {} 

\maketitle

\section{Introduction\label{intro}}

In the Galaxy, X-ray emissions are dominated by accreting compact objects and
isolated neutron stars (e.g. pulsars). Most black-hole X-ray binaries (BH XRBs)
are transients, and more than one hundred outbursts are recorded in the vast
database in approximately the past 20 years \citep[see, e.g.
][]{tetarenko15,corral16}. The phenomenology of the evolution of BH XRBs
outbursts appears complex, and different state classifications have been
proposed by different authors \citep[e.g. see the review by][]{remillard06,
zhang13}. Based on the hardness and timing properties, the spectral states of
BH XRBs can be distinguished into the low-hard state (LHS), the
hard-intermediate state (HIMS), the soft-intermediate state (SIMS), and the
high-soft state (HSS) \citep[see, e.g. ][]{belloni16}.

Relationships among the spectral and timing parameters are crucial to the study
of accretion flows in BH XRBs \citep[e.g.][]{belloni05, miller06, ingram11}.
Investigating the phase/time lag allows us to explore the connection between
the accretion disk and the corona and jet \citep[e.g.
][]{qu10,altamirano15,reig15,veledina15}. Three types of low frequency
quasi-periodic oscillations (QPOs), i.e. type-A, type-B, and type-C QPOs, have
been occasionally observed at different spectral states \citep[e.g.
][]{casella05,roy11,li13}. The study of low frequency QPOs provide an
opportunity to understand accretion flows close to central black-holes (BHs).
When sources transit from the HIMS to the SIMS, the type-C QPOs disappear and
the type-B QPOs might be present in the power density spectra (PDSs)
\citep[][and references therein]{belloni16}. It is worth noting that the
transition takes place with a minor change in hardness and sources cross the
so-called jet line in the hardness-intensity diagram (HID) \citep[e.g.
][]{fender04}. The association of the type-B QPOs and the ejection of
relativistic ballistic jets has been revealed in BH XRBs \citep[e.g.
][]{soleri08,kylafis15}. Various models taking instabilities and geometrical
effects into account, have been proposed to explain the origin and the
behaviour of QPOs in BH XRBs \citep[e.g.][]{stella98,ingram09,varniere12}.

The disk instability model (DIM) is widely accepted as the explanation for
outbursts in dwarf nova. The model has also been suggested to account for
outbursts in X-ray transients, but in the meantime various deficiencies have
been figured out \citep[see the review by][]{lasota01}. In order to reproduce
the major observational phenomena, the DIM should take the irradiation effect
and a transition to a radiatively inefficient accretion flow below a critical
mass-transfer rate into account \citep[e.g.][]{king98,coriat12}. Irradiation by
the central X-ray source is required to keep the outer disk hot and generate
slow exponential decay light curves \citep[e.g.][]{chen97,yan15}. This scenario
is supported by the correlation between the X-ray and the optical and
near-ultraviolet (NUV) luminosities \citep{van94,rykoff07}. Recently, Weng \&
Zhang (2015) analysed the multiwavelength light curve evolution of
\object{Swift j1357.2-0933} during its 2011 outburst, and found that the X-ray
reprocessing was negligible since the NUV luminosity was close to and even
exceeded the X-ray luminosity at the times; however, the light curves displayed
the typical near-exponential decay profile. This means that there is need for
improvement in the current version of DIM, and the origins of NUV emission are
diverse (i.e. stemming from the companion star, jet, the outer disk, etc).

\object{Swift J1842.5-1124} was discovered on July 2008 by the {\it Swift}/BAT
(Krimm et al. 2008). A series of {\it RXTE} and {\it Swift} observations were
carried out to follow up the outburst. The multi-band light curves had been
presented in Krimm et al. (2013), and the hard X-ray peak preceded the soft
X-ray peak by a few days, being consistent with those observed in other BH XRBs
\citep[e.g. ][]{zhou13}. In addition, both spectral and temporal fittings were
performed using a few individual observations \citep[e.g.][]{markwardt08}. But
there is still a lack of systematic study on the spectral and timing evolution
during the 2008 outburst. In this work, we take a detailed analysis on both
{\it RXTE} and {\it Swift} observations. The data reduction is described in
next section, and the results are presented in Section 3. Since \object{Swift
J1842.5-1124} showed some common observed properties with other BH XRBs, it was
recognized as a promising BH candidate. In Section 4, we discuss the analysis
results and compare them with other BH XRBs. The summary follows in Section 5.

\section{Observations and data reduction}

\begin{figure}%%%%%%%%%%%%%%%%%%%%%%%%%%%%%%%%%%%%%%%%%%%%%%%%%%%%%
\begin{center}
\centerline{\includegraphics[scale=0.5]{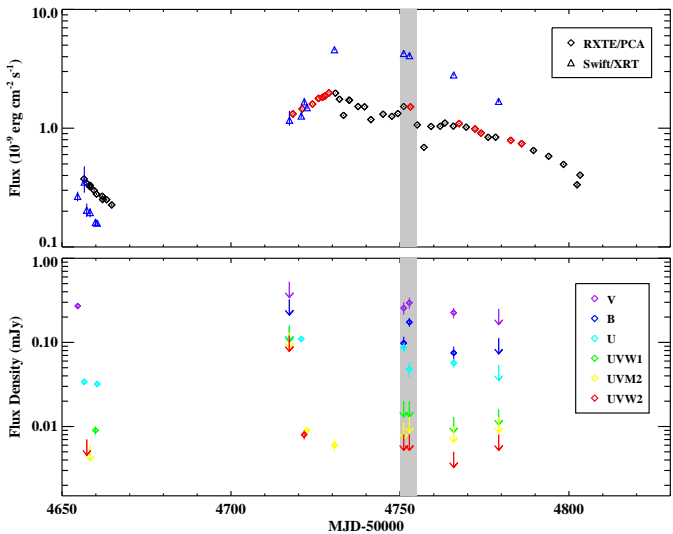}} \caption[]{X-ray (top
panel, Tables \ref{xrt} and \ref{rxte}) and optical and NUV light curves
(bottom panel) for the 2008 outburst of \object{Swift J1842.5-1124}. The red
symbols corresponds to the data in which QPOs were observed. Shaded areas
indicate the period when the source took the excursion in the HID (Figure
\ref{hid}).}\label{lc}
\end{center}
\end{figure}

\object{Swift J1842.5-1124} triggered the {\it Swift}/BAT in 2008 July (Krimm
et al. 2008), and was visited by {\it Swift} and {\it RXTE} on 16 and 49
occasions in 2008, respectively. The light curves for the activities are
displayed in Figure \ref{lc}.

\subsection{{\it Swift}}

The {\it Swift} Gamma Ray Burst Explorer carries three scientific instruments,
including the Burst Alter Telescope (BAT), the X-ray Telescope (XRT), and the
UV/Optical Telescope (UVOT) \citep{gehrels04}. Because of broadband coverage
and scheduling flexibility, {\it Swift} is ideally suited to tracing outbursts
of accreting X-ray binaries. Both the XRT and UVOT data are processed with the
packages and tools available in HEASOFT version 6.17 in the standard way. The
XRT operating mode switched between the photon counting (PC) and the windowed
timing (WT) modes in order to minimize the pile-up effect. The raw data are
performed for the initial event cleaning using the \texttt{xrtpipeline} script.
The source spectra are extracted within a circle of radius 20 pixels centred at
the nominal position of \object{Swift J1842.5-1124} (R.A. = 18:42:17.45, decl.
= -11:25:03.9, J2000) with \texttt{xselect}, while an annulus region with the
radius of 20 and 40 pixels is adopted for background region. The ancillary
response files are produced with task \texttt{xrtmkarf}, and the latest
response files (v014) are taken from the CALDB database. We also rebin the
spectra to have at least 20 counts per bin to enable the use of $\chi^{2}$
statistics. Because there are calibration residuals at the energy below 0.6 keV
for the WT mode data \footnote{see
http://swift.gsfc.nasa.gov/docs/heasarc/caldb/swift/docs/xrt/
SWIFTXRT-CALDB-09.pdf}, the spectral fitting are restricted to the 0.6--10 keV
band.

For the X-ray spectral analysis in this work, we begin by fitting an absorbed
power-law model to the spectra, and then a multicolour disk blackbody model is
added if it has a significance level above 99\% based on the $F-$test (Table
\ref{xrt}). Coincidentally, all observations performed in the PC mode have
fewer photons, which can be fitted by the single power-law model. On the other
hand, all WT mode data have high signal-to-noise ratios, and an additional disk
blackbody component is required. Since we do not expect the absorption column
density $N_{\rm H}$ to be different among the {\it Swift} observations, we also
try to fit all PC mode data simultaneously with the same value of $N_{\rm H}$,
but leave the power-law model parameters untied, yielding $N_{\rm H} =
(2.7\pm0.5)\times10^{21} {\rm cm}^{-2}$. Alternatively, we obtain $N_{\rm H} =
(4.2\pm0.1)\times10^{21} {\rm cm}^{-2}$ if modelling all WT mode observations
with a common value of $N_{\rm H}$. It is worth to note that the value of
$N_{\rm H}$ varies with adopted models; however, our results shown below are
not dependent on the precise value of $N_{\rm H}$.

%%%%%%%%%%%%%%%%%%%%%%%%%%%%%%%%

\begin{table*}
\begin{center}
\begin{tabular}{c c c c c c c c}
\hline
Obs.ID   &  MJD    &      $N_{\rm H}$     &  $kT$  &  $\Gamma$   &  Flux$_{\rm Po}$  & Flux$_{\rm Total}$ & $\chi^2$/dof \\
        &          & ($10^{22}$ cm$^{-2}$)  &  (keV)   &             &   ($10^{-9}$\ferg)  &   ($10^{-9}$\ferg)   &         \\ \hline
00031234001$^{\S}$& 54654.6 & $0.30_{-0.14}^{+0.16}$ &          --            & $1.50_{-0.22}^{+0.24}$ & $0.27_{-0.02}^{+0.02}$ & $0.27_{-0.02}^{+0.02}$ & 25.0/36 \\
00031234002& 54656.6 & $0.55_{-0.17}^{+0.16}$ & $0.12_{-0.01}^{+0.02}$ & $1.64_{-0.11}^{+0.11}$ & $0.28_{-0.01}^{+0.01}$ & $0.35_{-0.07}^{+0.12}$ & 217.6/249\\
00031234003$^{\S}$& 54657.3 & $0.53_{-0.20}^{+0.23}$ &          --            & $1.75_{-0.33}^{+0.37}$ & $0.20_{-0.02}^{+0.03}$ & $0.20_{-0.02}^{+0.03}$ & 14.9/19 \\
00031234004$^{\S}$& 54658.3 & $0.26_{-0.11}^{+0.12}$ &          --            & $1.24_{-0.20}^{+0.22}$ & $0.19_{-0.02}^{+0.02}$ & $0.19_{-0.02}^{+0.02}$ & 42.2/41 \\
00031234005$^{\S}$& 54659.9 & $0.26_{-0.10}^{+0.11}$ &          --            & $1.33_{-0.17}^{+0.18}$ & $0.16_{-0.01}^{+0.01}$ & $0.16_{-0.01}^{+0.01}$ & 48.7/52 \\
00031234006$^{\S}$& 54660.4 & $0.22_{-0.07}^{+0.08}$ &          --            & $1.23_{-0.13}^{+0.14}$ & $0.16_{-0.01}^{+0.01}$ & $0.16_{-0.01}^{+0.01}$ & 77.4/74 \\
00324112000$^{\sharp}$& 54717.3 & $0.33_{-0.10}^{+0.13}$ & $0.22_{-0.05}^{+0.14}$ & $1.68_{-0.13}^{+0.11}$ & $1.02_{-0.05}^{+0.05}$ & $1.16_{-0.11}^{+0.23}$ & 198.0/187\\
00031234007& 54720.8 & $0.29_{-0.04}^{+0.08}$ & $0.34_{-0.12}^{+0.14}$ & $1.70_{-0.13}^{+0.11}$ & $1.16_{-0.08}^{+0.06}$ & $1.27_{-0.05}^{+0.12}$ & 355.6/301\\
00031234008& 54721.7 & $0.36_{-0.04}^{+0.06}$ & $0.24_{-0.05}^{+0.07}$ & $1.92_{-0.06}^{+0.06}$ & $1.50_{-0.05}^{+0.05}$ & $1.67_{-0.08}^{+0.13}$ & 434.0/398\\
00031234009& 54722.5 & $0.30_{-0.02}^{+0.04}$ & $0.37_{-0.09}^{+0.09}$ & $1.75_{-0.10}^{+0.09}$ & $1.34_{-0.07}^{+0.06}$ & $1.49_{-0.04}^{+0.06}$ & 321.5/324\\
00031234010& 54730.6 & $0.39_{-0.01}^{+0.02}$ & $0.74_{-0.01}^{+0.01}$ & $1.83_{-0.15}^{+0.13}$ & $1.74_{-0.22}^{+0.24}$ & $4.58_{-0.07}^{+0.09}$ & 597.1/552\\
00031234011& 54751.1 & $0.45_{-0.01}^{+0.01}$ & $0.67_{-0.01}^{+0.01}$ & $1.77_{-0.14}^{+0.13}$ & $1.67_{-0.18}^{+0.21}$ & $4.26_{-0.06}^{+0.07}$ & 760.4/525\\
00031234012& 54752.8 & $0.43_{-0.02}^{+0.02}$ & $0.64_{-0.01}^{+0.01}$ & $2.03_{-0.14}^{+0.13}$ & $1.96_{-0.25}^{+0.28}$ & $4.09_{-0.10}^{+0.11}$ & 574.5/496\\
00031234013& 54765.9 & $0.40_{-0.01}^{+0.01}$ & $0.60_{-0.01}^{+0.01}$ & $2.02_{-0.11}^{+0.10}$ & $1.37_{-0.13}^{+0.14}$ & $2.81_{-0.05}^{+0.06}$ & 531.1/543\\
00031234014& 54779.2 & $0.39_{-0.02}^{+0.02}$ & $0.45_{-0.03}^{+0.02}$ & $2.11_{-0.11}^{+0.10}$ & $1.22_{-0.11}^{+0.11}$ & $1.68_{-0.04}^{+0.05}$ & 497.4/418\\

\hline
\end{tabular}
\caption{Best fit results for {\it Swift}/XRT observations. Flux$_{\rm Po}$:
0.6--10 keV unabsorbed flux for the power-law component. Flux$_{\rm Total}$:
total unabsorbed flux calculated in 0.6--10 keV.  $^{\S}$: Observations were
operated in the photon-counting mode, and the rest of {\it Swift}/XRT
observations were executed in the window-timing mode. $^{\sharp}$: There were
two short automated observations (Obs IDs = 00324112000 and 00324116000)
performed within two hours on 2008 September 14 (MJD = 54717); hence we fit
them together in order to improve the signal-to-noise. \label{xrt}}
\end{center}
\end{table*}

%%%%%%%%%%%%%%%%%%%%%%%%%%%%%%%%%%%%%%%

The UVOT data were taken in the image mode. In order to increase photon
statistics, we stack the images when there is more than one exposure in the
observations by using \texttt{uvotimsum}. An aperture radius 5 arcsec is
adopted for the aperture photometry in the stacked images, and the background
flux density is measured from a neighbouring source free sky region. We also
estimate the 3$\sigma$ upper limits if the source was undetected.

\subsection{{\it RXTE}}

%%%%%%%%%%%%%%%%%%%%%%%%%%%%%%%%%%%%%%%%%%%%%%%%%%%%%
\begin{figure}
\begin{center}
\centerline{\includegraphics[scale=0.38]{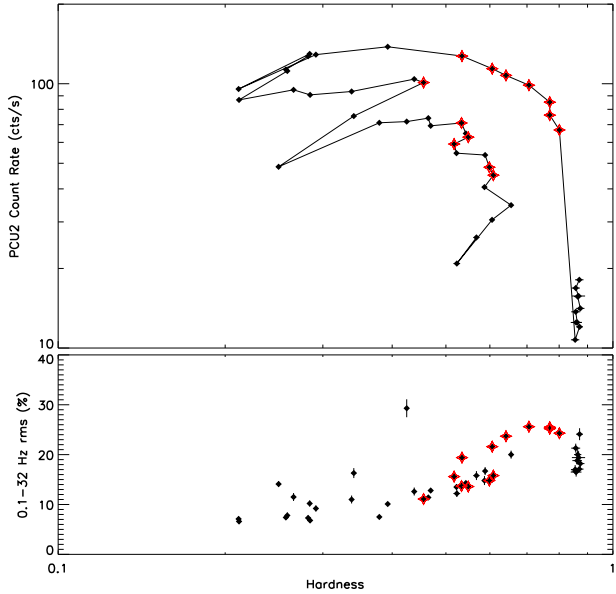}}
\caption[]{Hardness-intensity diagram (HID: top panel) and hardness-rms diagram
(HRD: bottom panel) for the 2008 outburst of \object{Swift J1842.5-1124}. The
red symbols corresponds to the data in which QPOs were detected.}\label{hid}
\end{center}
\end{figure}

%The caveat is that we do not discuss the QPOs observed during the source return to the LHS because of low quality data.
%%%%%%%%%%%%%%%%%%%%%%%%%%%%%%%%%%%%%%%%%%%%%

The denser {\it RXTE} observations were executed to cover the activities of
\object{Swift J1842.5-1124}, especially from 2008 September to November. In
this work, we focus on data from the main instrument of {\it RXTE} --- the
Proportional Counter Array (PCA). The data from the PCU2 are filtered with the
standard criteria: the Earth-limb elevation angle larger than $10\degr$ and the
spacecraft pointing offset less than $0.02\degr$. The bright and faint
background models are used for creating background files when the source
intensities were larger and lower than 40 count s$^{-1}$ PCU$^{-1}$,
respectively. We produce the background-subtracted light curves in the PCA
channels of 8-14 (3.7--6.3 keV) and 15-25 (6.3--10.6 keV), then average their
count rates in each observation and plot the HID in Figure \ref{hid}. We first
fit the PCA spectra with an absorbed power-law model, and then include an
additional multicolour disk blackbody model when its significance is larger
than 99\%, as we did to the {\it Swift/XRT} data, while the neutral hydrogen
column density in the PCA spectral fitting is fixed to $3.6\times10^{21}$
cm$^{-2}$ according to the {\it Swift}/XRT spectral fitting (Table \ref{xrt}).
An absorption edge around 7 keV is also added to account for the line feature,
and the recommended systematic error of 0.5\% is applied. The fitting results
are shown in Table \ref{rxte}.

\begin{table*}
\begin{center}
\begin{tabular}{c c c c c c c}
\hline
Obs.ID   &  MJD    &  $kT$  &  $\Gamma$   &  Flux$_{\rm Po}$  & Flux$_{\rm Total}$ & $\chi^2$/dof \\
        &          &  (keV)   &             &   ($10^{-9}$\ferg)  &   ($10^{-9}$\ferg)   &         \\ \hline
93065-04-01-00&      54656.5&          --             & $1.55_{-0.03}^{+0.03}$ & $0.37_{-0.01}^{+0.01}$ & $0.37_{-0.01}^{+0.01}$ & 26.3/45 \\
93065-04-01-01&      54657.4&          --             & $1.59_{-0.03}^{+0.03}$ & $0.34_{-0.01}^{+0.01}$ & $0.34_{-0.01}^{+0.01}$ & 38.9/45 \\
93065-04-02-00&      54658.2&          --             & $1.60_{-0.04}^{+0.05}$ & $0.32_{-0.01}^{+0.01}$ & $0.32_{-0.01}^{+0.01}$ & 35.4/45 \\
93065-04-02-01&      54658.3&          --             & $1.57_{-0.03}^{+0.03}$ & $0.33_{-0.01}^{+0.01}$ & $0.33_{-0.01}^{+0.01}$ & 32.8/45 \\
93065-04-02-02&      54659.5&          --             & $1.57_{-0.03}^{+0.03}$ & $0.30_{-0.01}^{+0.01}$ & $0.30_{-0.01}^{+0.01}$ & 41.2/45 \\
93065-04-02-03&      54660.2&          --             & $1.57_{-0.03}^{+0.04}$ & $0.28_{-0.01}^{+0.01}$ & $0.28_{-0.01}^{+0.01}$ & 32.9/45 \\
93065-04-02-04&      54661.9&          --             & $1.57_{-0.04}^{+0.04}$ & $0.27_{-0.01}^{+0.01}$ & $0.27_{-0.01}^{+0.01}$ & 30.5/45 \\
93065-04-02-05&      54662.0&          --             & $1.61_{-0.05}^{+0.05}$ & $0.25_{-0.01}^{+0.01}$ & $0.25_{-0.01}^{+0.01}$ & 39.7/45 \\
93065-04-02-06&      54663.2&          --             & $1.60_{-0.03}^{+0.03}$ & $0.25_{-0.01}^{+0.01}$ & $0.25_{-0.01}^{+0.01}$ & 34.0/45 \\
93065-04-02-07&      54664.7&          --             & $1.59_{-0.03}^{+0.03}$ & $0.23_{-0.01}^{+0.01}$ & $0.23_{-0.01}^{+0.01}$ & 34.7/45 \\
93065-04-03-00&      54718.3&  $1.59_{-0.14}^{+0.11}$ & $1.51_{-0.06}^{+0.06}$ & $1.20_{-0.04}^{+0.04}$ & $1.32_{-0.01}^{+0.01}$ & 51.5/43 \\
93065-04-04-00&      54721.1&  $1.60_{-0.14}^{+0.10}$ & $1.54_{-0.08}^{+0.07}$ & $1.23_{-0.04}^{+0.04}$ & $1.45_{-0.01}^{+0.01}$ & 49.9/43 \\
93065-04-04-01&      54724.1&  $1.61_{-0.20}^{+0.13}$ & $1.64_{-0.09}^{+0.08}$ & $1.49_{-0.08}^{+0.08}$ & $1.59_{-0.01}^{+0.01}$ & 51.5/43 \\
93065-04-04-02&      54725.9&  $1.35_{-0.19}^{+0.12}$ & $1.77_{-0.07}^{+0.06}$ & $1.65_{-0.07}^{+0.06}$ & $1.78_{-0.01}^{+0.01}$ & 62.7/43 \\
93065-04-04-04&      54727.1&  $1.17_{-0.40}^{+0.17}$ & $1.95_{-0.08}^{+0.07}$ & $1.70_{-0.09}^{+0.07}$ & $1.82_{-0.01}^{+0.01}$ & 48.7/43 \\
93065-04-04-03&      54727.8&  $1.03_{-0.20}^{+0.14}$ & $2.04_{-0.05}^{+0.04}$ & $1.74_{-0.06}^{+0.05}$ & $1.86_{-0.01}^{+0.01}$ & 60.5/43 \\
93065-04-06-00&      54729.0&  $0.72_{-0.09}^{+0.09}$ & $2.19_{-0.03}^{+0.03}$ & $1.84_{-0.03}^{+0.02}$ & $1.98_{-0.01}^{+0.01}$ & 67.4/43 \\
93065-04-06-01&      54730.9&  $0.70_{-0.03}^{+0.03}$ & $2.35_{-0.04}^{+0.04}$ & $1.58_{-0.02}^{+0.02}$ & $1.97_{-0.01}^{+0.01}$ & 53.7/43 \\
93065-04-06-02&      54732.1&  $0.69_{-0.01}^{+0.01}$ & $2.40_{-0.05}^{+0.05}$ & $1.15_{-0.02}^{+0.02}$ & $1.75_{-0.01}^{+0.01}$ & 56.3/43 \\
93065-04-06-04&      54733.3&  $0.69_{-0.01}^{+0.01}$ & $2.12_{-0.09}^{+0.09}$ & $0.59_{-0.01}^{+0.02}$ & $1.28_{-0.01}^{+0.01}$ & 64.8/43 \\
93065-04-06-03&      54735.0&  $0.72_{-0.02}^{+0.02}$ & $2.26_{-0.09}^{+0.08}$ & $1.08_{-0.03}^{+0.03}$ & $1.71_{-0.01}^{+0.01}$ & 40.1/43 \\
93065-04-05-00&      54735.0&  $0.71_{-0.02}^{+0.01}$ & $2.31_{-0.07}^{+0.07}$ & $1.06_{-0.02}^{+0.02}$ & $1.72_{-0.01}^{+0.01}$ & 46.1/43 \\
93065-04-05-01&      54737.6&  $0.70_{-0.01}^{+0.01}$ & $2.30_{-0.06}^{+0.06}$ & $0.86_{-0.01}^{+0.02}$ & $1.52_{-0.01}^{+0.01}$ & 50.2/43 \\
93111-01-01-00&      54739.5&  $0.67_{-0.01}^{+0.01}$ & $2.35_{-0.04}^{+0.05}$ & $0.89_{-0.01}^{+0.01}$ & $1.51_{-0.01}^{+0.01}$ & 54.9/43 \\
93111-01-01-01&      54741.4&  $0.68_{-0.01}^{+0.01}$ & $2.10_{-0.06}^{+0.07}$ & $0.56_{-0.01}^{+0.01}$ & $1.18_{-0.01}^{+0.01}$ & 50.3/43 \\
93111-01-02-00&      54745.0&  $0.67_{-0.01}^{+0.02}$ & $2.22_{-0.09}^{+0.06}$ & $0.76_{-0.02}^{+0.01}$ & $1.31_{-0.01}^{+0.01}$ & 65.1/43 \\
93111-01-02-02&      54747.6&  $0.65_{-0.01}^{+0.01}$ & $2.30_{-0.06}^{+0.06}$ & $0.79_{-0.01}^{+0.01}$ & $1.26_{-0.01}^{+0.01}$ & 58.0/43 \\
93111-01-03-00&      54749.4&  $0.67_{-0.02}^{+0.02}$ & $2.27_{-0.05}^{+0.05}$ & $0.94_{-0.01}^{+0.01}$ & $1.32_{-0.01}^{+0.01}$ & 52.9/43 \\
93111-01-03-01&      54751.1&  $0.66_{-0.04}^{+0.04}$ & $2.32_{-0.05}^{+0.05}$ & $1.29_{-0.02}^{+0.02}$ & $1.52_{-0.01}^{+0.01}$ & 40.1/43 \\
93111-01-03-02&      54753.1&  $0.64_{-0.04}^{+0.04}$ & $2.28_{-0.04}^{+0.03}$ & $1.31_{-0.02}^{+0.02}$ & $1.51_{-0.01}^{+0.01}$ & 83.3/43 \\
93111-01-03-03&      54755.1&  $0.58_{-0.03}^{+0.03}$ & $2.43_{-0.08}^{+0.08}$ & $0.79_{-0.01}^{+0.01}$ & $1.06_{-0.01}^{+0.01}$ & 27.4/43 \\
93111-01-04-00&      54757.1&  $0.61_{-0.01}^{+0.01}$ & $2.14_{-0.06}^{+0.06}$ & $0.38_{-0.01}^{+0.01}$ & $0.69_{-0.01}^{+0.01}$ & 36.2/43 \\
93111-01-04-01&      54759.2&  $0.62_{-0.02}^{+0.02}$ & $2.30_{-0.04}^{+0.04}$ & $0.80_{-0.01}^{+0.01}$ & $1.03_{-0.01}^{+0.01}$ & 40.6/43 \\
93111-01-04-03&      54761.9&  $0.53_{-0.04}^{+0.05}$ & $2.45_{-0.06}^{+0.05}$ & $0.89_{-0.01}^{+0.01}$ & $1.03_{-0.01}^{+0.01}$ & 26.8/43 \\
93111-01-05-00&      54763.3&  $0.56_{-0.04}^{+0.04}$ & $2.34_{-0.03}^{+0.03}$ & $0.99_{-0.01}^{+0.01}$ & $1.10_{-0.01}^{+0.01}$ & 32.3/43 \\
93111-01-05-01&      54765.8&  $0.53_{-0.04}^{+0.04}$ & $2.33_{-0.04}^{+0.04}$ & $0.93_{-0.01}^{+0.01}$ & $1.04_{-0.01}^{+0.01}$ & 38.0/43 \\
93111-01-05-02&      54767.5&  $0.50_{-0.07}^{+0.08}$ & $2.28_{-0.04}^{+0.03}$ & $1.03_{-0.01}^{+0.01}$ & $1.09_{-0.01}^{+0.01}$ & 39.2/43 \\
93111-01-05-03&      54769.6&  $0.60_{-0.12}^{+0.12}$ & $2.20_{-0.05}^{+0.05}$ & $0.96_{-0.02}^{+0.01}$ & $1.02_{-0.01}^{+0.01}$ & 43.4/43 \\
93111-01-06-00&      54772.2&  $0.47_{-0.06}^{+0.06}$ & $2.23_{-0.02}^{+0.02}$ & $0.93_{-0.01}^{+0.01}$ & $0.98_{-0.01}^{+0.01}$ & 37.8/43 \\
93111-01-06-01&      54774.0&  $0.57_{-0.08}^{+0.08}$ & $2.25_{-0.06}^{+0.06}$ & $0.83_{-0.01}^{+0.01}$ & $0.90_{-0.01}^{+0.01}$ & 54.0/43 \\
93111-01-06-02&      54776.1&  $0.49_{-0.07}^{+0.07}$ & $2.27_{-0.05}^{+0.04}$ & $0.78_{-0.01}^{+0.01}$ & $0.84_{-0.01}^{+0.01}$ & 25.6/43 \\
93111-01-07-00&      54778.3&  $0.42_{-0.11}^{+0.12}$ & $2.17_{-0.03}^{+0.03}$ & $0.83_{-0.01}^{+0.01}$ & $0.83_{-0.01}^{+0.01}$ & 24.9/43 \\
93111-01-07-02&      54782.8&  $0.33_{-0.11}^{+0.12}$ & $2.15_{-0.03}^{+0.03}$ & $0.76_{-0.01}^{+0.01}$ & $0.79_{-0.01}^{+0.01}$ & 40.7/43 \\
93111-01-08-00&      54786.0&  $0.57_{-0.19}^{+0.21}$ & $2.11_{-0.04}^{+0.03}$ & $0.72_{-0.01}^{+0.01}$ & $0.74_{-0.01}^{+0.01}$ & 34.2/43 \\
93111-01-08-01&      54789.5&  $0.43_{-0.12}^{+0.13}$ & $2.18_{-0.05}^{+0.05}$ & $0.62_{-0.01}^{+0.01}$ & $0.64_{-0.01}^{+0.01}$ & 24.9/43 \\
93454-01-01-00&      54794.0&  $0.76_{-0.43}^{+0.36}$ & $2.00_{-0.07}^{+0.05}$ & $0.57_{-0.02}^{+0.01}$ & $0.58_{-0.01}^{+0.01}$ & 25.1/43 \\
93454-01-02-00&      54798.4&  $0.22_{-0.16}^{+0.09}$ & $2.17_{-0.04}^{+0.03}$ & $0.48_{-0.01}^{+0.01}$ & $0.49_{-0.01}^{+0.01}$ & 36.9/43 \\
93454-01-02-01&      54802.4&  $0.55_{-0.09}^{+0.09}$ & $2.10_{-0.11}^{+0.10}$ & $0.29_{-0.01}^{+0.01}$ & $0.33_{-0.01}^{+0.01}$ & 27.3/43 \\
93454-01-02-02&      54803.3&  $0.53_{-0.13}^{+0.14}$ & $2.15_{-0.08}^{+0.07}$ & $0.39_{-0.01}^{+0.01}$ & $0.40_{-0.01}^{+0.01}$ & 34.8/43 \\

\hline
\end{tabular}
\caption{Best fit results for {\it RXTE}/PCA data with the absorption column
density fixed to $3.6\times10^{21}$ cm$^{-2}$. \label{rxte}}
\end{center}
\end{table*}

%%%%%%%%%%%%%%%%%%%%%%%%%%%%%%%%%%%%%%%%%%%%%%%

The light curves in the channel of 0-43 (2--19 keV) are extracted from the PCA
Event mode data, E\_125us\_64M\_0\_1s for temporal analysis. The data are
divided into 32 s segments with 8 ms time bins, and the PDSs are generated with
the task \texttt{powspec}. We adopt the Miyamoto method to normalize the PDSs
\citep{miyamoto91}, and average them using a logarithmic rebinning. After
subtracting the Poisson noise, we integrate the fractional root mean square
(rms) in the 0.1--32 Hz, and plot the hardness-rms diagram (HRD) in the bottom
panel of Figure \ref{hid}. However, the low signal-to-noise ratio of data do
not allow us to carry out energy-dependent studies, for example the energy
dependence of the centroid frequency and phase lag of QPOs.

\begin{figure}%%%%%%%%%%%%%%%%%%%%%%%%%%%%%%%%%%%%%%%%%%%%%%%%%%%%%
\begin{center}
\centerline{\includegraphics[scale=0.65]{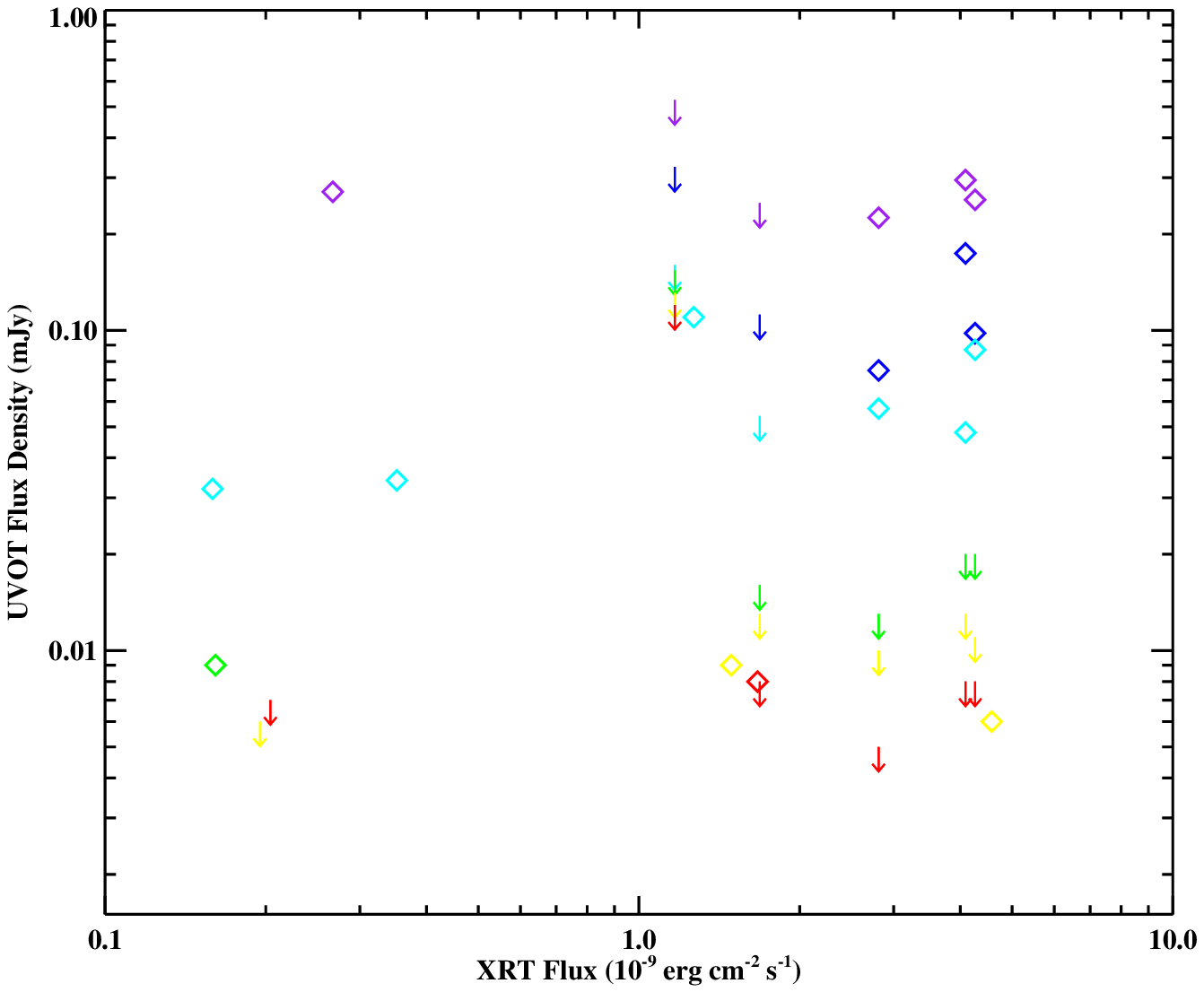}} \caption[]{Correlation
between the optical/NUV flux density and the X-ray 0.6--10 keV flux. The
meaning of different colours is the same as in the bottom panel of Figure
\ref{lc}. We do not plot the errors of data for clarify; while readers can find
them in Figure \ref{lc}.}\label{lxluv}
\end{center}
\end{figure}

%%%%%%%%%%%%%%%%%%%%%%%%%%%%%%%%%%%%%%%%%%%%%%%%%%%%%
\begin{figure*}
\begin{center}
\centerline{\includegraphics[scale=0.4]{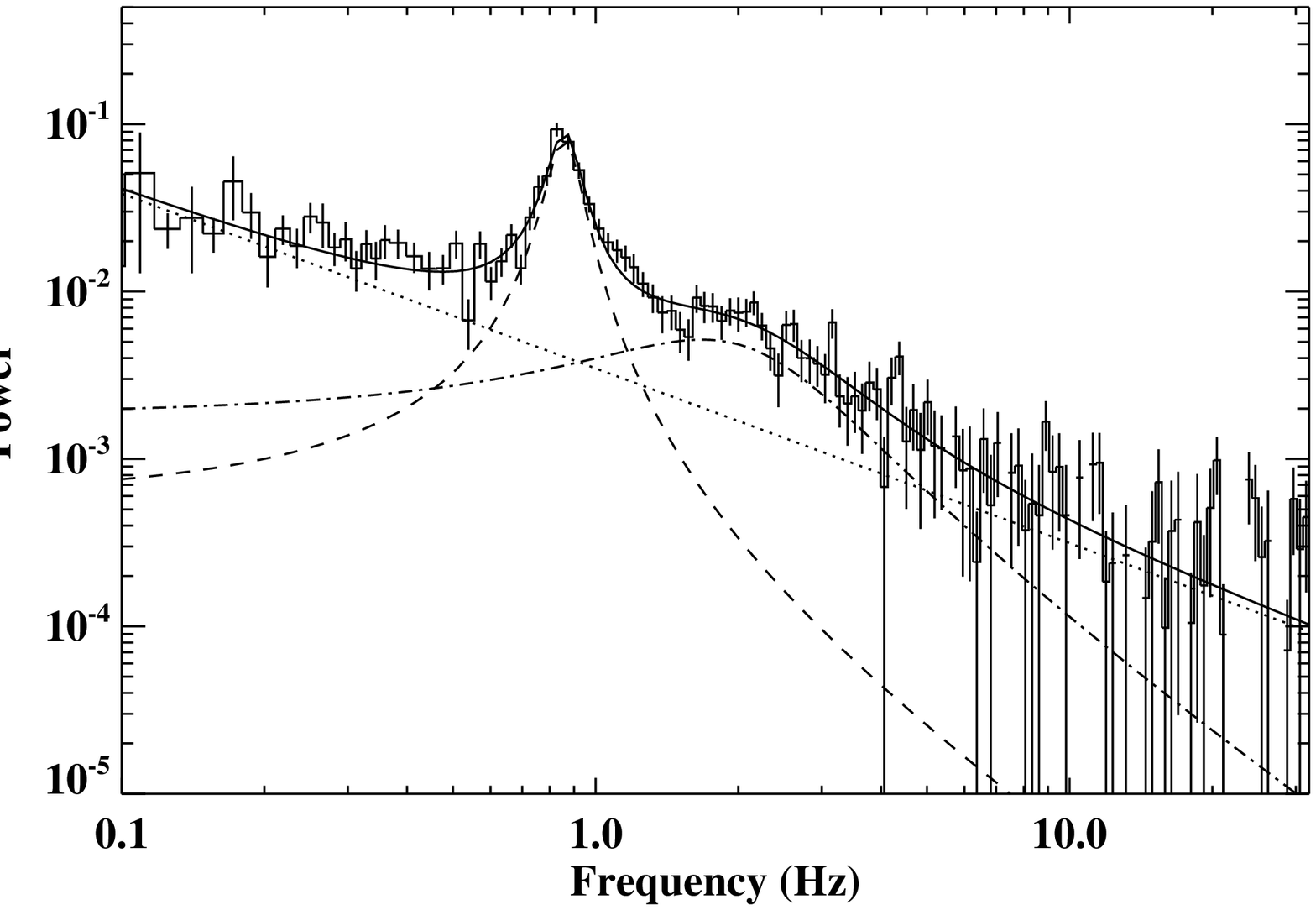}
\includegraphics[scale=0.4]{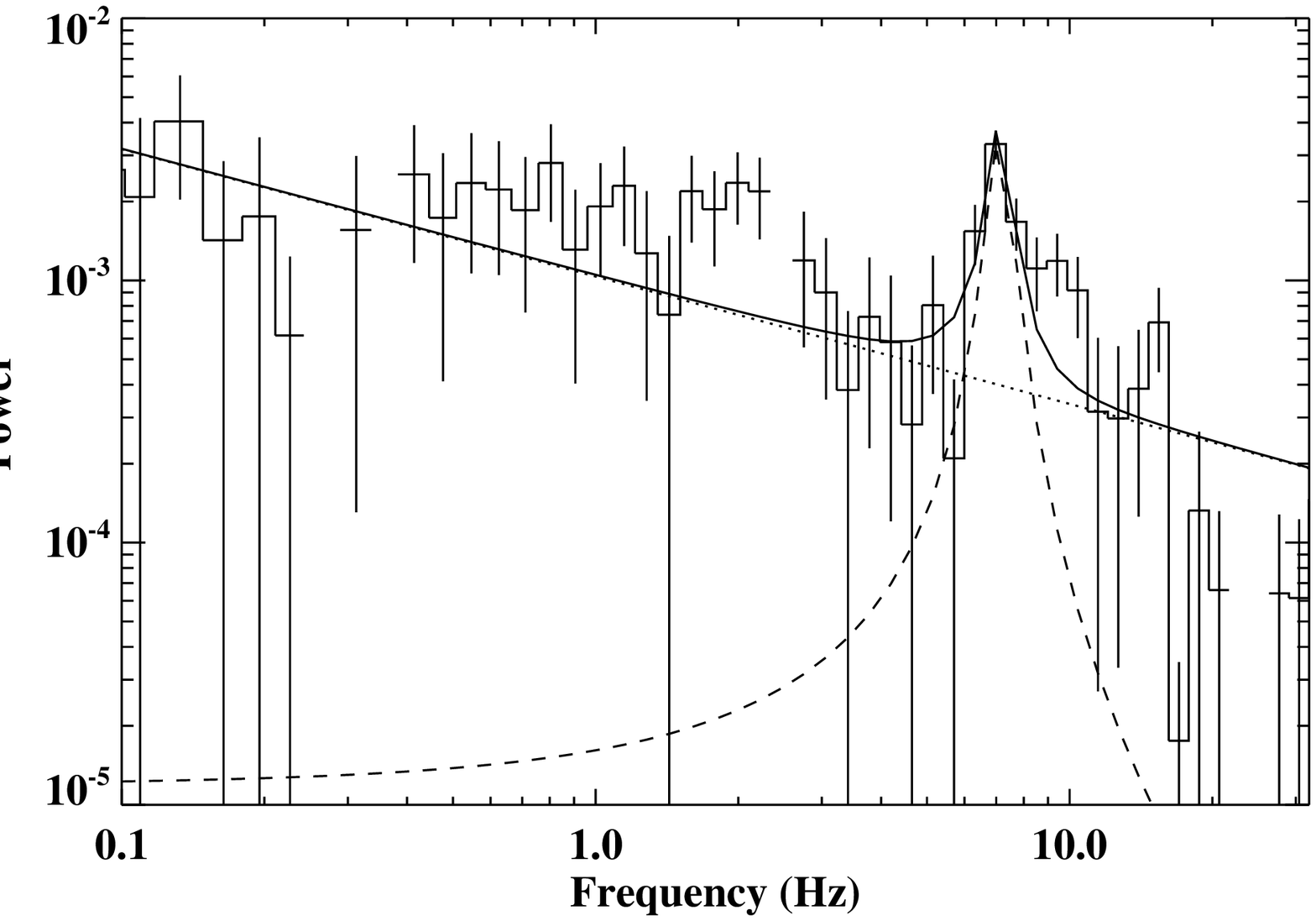}}
\caption[]{Left panel: PDS for MJD 54718.3 (Obs ID = 93065-04-03-00) is fitted
by a power-law plus two Lorentzian models. Right panel: PDS for MJD 54772.2
(Obs ID = 93111-01-06-00) is fitted by a power-law model with an additional
Lorentzian component. We note that the PDS is rebinned for
clarity.}\label{power}
\end{center}
\end{figure*}

%%%%%%%%%%%%%%%%%%%%%%%%%%%%%%%%%%%%%%%%%%%%%

\begin{table}
\begin{center}
\begin{tabular}{c c c c }
\hline Obs.ID   &  Frequency (Hz)    &      FWHM (Hz)    &  rms$_{\rm QPO}$
\\\hline

%93065-04-02-01&  $0.17_{-0.01}^{+0.01}$ & 7.80& 7.66$\pm$7.66 \\
%93065-04-02-02&  $0.14_{-0.01}^{+0.01}$ & 3.53& 8.39$\pm$5.54 \\
LHS to HSS &&& \\
93065-04-03-00&  $0.85_{-0.01}^{+0.01}$ & $0.14_{-0.02}^{+0.02}$& 19.44$\pm$3.13 \\
93065-04-04-00&  $1.15_{-0.01}^{+0.01}$ & $0.21_{-0.03}^{+0.03}$& 21.27$\pm$3.06 \\
93065-04-04-01&  $1.36_{-0.01}^{+0.01}$ & $0.22_{-0.03}^{+0.03}$& 21.20$\pm$2.84 \\
93065-04-04-02&  $2.17_{-0.02}^{+0.02}$ & $0.32_{-0.04}^{+0.05}$& 21.80$\pm$3.17 \\
93065-04-04-04&  $3.07_{-0.02}^{+0.02}$ & $0.35_{-0.04}^{+0.05}$& 19.66$\pm$2.70 \\
93065-04-04-03&  $3.75_{-0.03}^{+0.03}$ & $0.44_{-0.06}^{+0.07}$& 18.17$\pm$2.86 \\
93065-04-06-00&  $5.31_{-0.03}^{+0.03}$ & $0.58_{-0.06}^{+0.07}$& 14.68$\pm$1.71 \\
\hline
HSS to LHS &&& \\
93111-01-03-02&  $7.98_{-0.39}^{+0.31}$ & $1.45_{-0.83}^{+1.55}$& 8.16$\pm$6.90 \\
93111-01-05-02&  $7.18_{-0.30}^{+0.27}$ & $1.71_{-0.80}^{+1.33}$& 12.12$\pm$7.25 \\
93111-01-06-00&  $7.12_{-0.10}^{+0.11}$ & $0.81_{-0.29}^{+0.51}$& 9.70$\pm$4.68 \\
93111-01-06-01&  $8.18_{-0.25}^{+0.35}$ & $0.86_{-0.57}^{+0.96}$& 8.80$\pm$8.69 \\
93111-01-07-02&  $6.13_{-0.20}^{+0.18}$ & $0.79_{-0.40}^{+0.80}$& 9.41$\pm$7.08 \\
93111-01-08-00&  $5.56_{-0.19}^{+0.16}$ & $1.00_{-0.47}^{+0.77}$& 11.14$\pm$6.68 \\

\hline
\end{tabular}
\caption{Log of QPOs. \label{qpo_log}}
\end{center}
\end{table}

%%%%%%%%%%%%%%%%%%%%%%%%%%%%%%%%%%%%%%%%%%%%%%%%%%%%%
\begin{figure}
\begin{center}
\centerline{\includegraphics[scale=0.38]{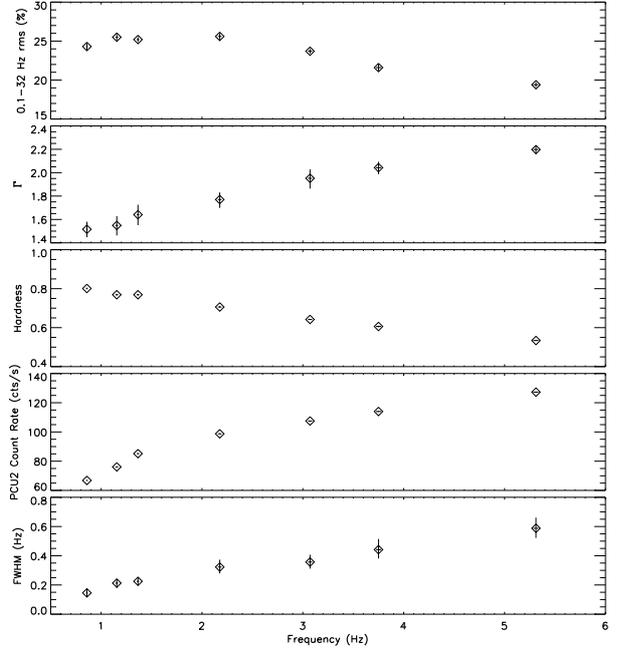}} \caption[]{Panels from top
to bottom show the total fractional rms, photon index $\Gamma$, hardness,
intensity, and the FWHM plotted against the QPO frequency,
respectively.}\label{qpo}
\end{center}
\end{figure}

%%%%%%%%%%%%%%%%%%%%%%%%%%%%%

\section{Results}

Since the first X-ray observation was taken in 2008 July, the X-ray flux of
\object{Swift J1842.5-1124} declined exponentially with time and became
undetectable two weeks later. After a few days, the source re-brightened and
reached the X-ray peak around the end of September (MJD $\sim$ 54730, Figure
\ref{lc}). At the peak of outburst, the 0.6--10 keV X-ray flux estimated from
the {\it Swift}/XRT observations were even higher than the 3-25 keV flux
measured by {\it RXTE}/PCA, that is the X-ray spectra were relatively softer
(Tables \ref{xrt} and \ref{rxte}). However, the source was sparsely detected in
the NUV bands (UW1, UM2, and UW2) around the peak of outburst. The flux density
measured by the V filter remained nearly constant $\sim 0.26$ mJy (with a
slightly increase on MJD 54753) as the X-ray luminosity varied. Alternatively,
the source was observed in the U band six times, and there was a weakly
positive correlation between the U-band flux density and the X-ray flux with
the Spearman's rank correlation coefficient of $\rho/P = 0.60/0.21$. In Figure
\ref{lxluv}, we plot the optical/NUV flux density versus the corresponding
X-ray flux detected by {\it Swift}. Using the relation between the optical
extinction and the hydrogen column density given by \cite{guver09}, we find
that the optical and NUV flux measured by each filter ($\sim 10^{-12}$ erg
cm$^{-2}$ s$^{-1}$) is smaller than the X-ray flux by about three orders of
magnitude.

In the HID, \object{Swift J1842.5-1124} displayed a `q' shaped diagram,
travelling counterclockwise from the bottom right corner with an additional
excursion to a harder state and back in 2008 October (Figure \ref{hid}). On the
other hand, the total fractional rms generally decreased with the increasing
hardness, and no hysteresis was observed in the HRD. That is, both the HID and
HRD of \object{Swift J1842.5-1124} resembled those found in other BH XRBs
\citep[e.g.][]{dunn10,motta11,zhou13}. Therefore, following the state
definition described in Belloni \& Motta (2016), we suggest that \object{Swift
J1842.5-1124} went through the LHS (the bottom right corner in the HID), the
HIMS (the upper right corner), the HSS (the upper left corner), and then
returned the LHS via the SIMS/HIMS during its 2008 outburst. In addition to the
hardness and timing properties shown in the HID and HRD, the state
identifications are supported by more detailed spectral fitting results (Tables
\ref{xrt} and \ref{rxte}) and timing analysis. The harder X-ray spectra
accompanied by the strong aperiodic variabilities were observed in the LHS,
while the larger values of photon index and weaker variabilities were found
when the source transited to the HSS. Compared to the transition from the LHS
to the HSS, the transition from the HSS to the LHS exhibited softer spectra
with a photon index $\Gamma \sim 2.0-2.2$ (Table \ref{rxte}), that is the
hysteresis.

A strong QPO with a frequency of $\sim 0.8$ Hz had been reported by
\cite{markwardt08}, and a weak QPO at $\sim 8$ Hz presented on 2008 October 14
(MJD = 54753) was also suggested \citep{krimm13}. In this paper, we employ the
power-law and Lorentzian models in the PDSs fitting. The strong QPOs were
revealed in the seven observations when the source transited from the LHS to
the HSS; while the weak QPOs were displayed in the PDSs of another six
observations during the source returned to LHS (Figure \ref{hid}). As examples
for both cases, we plot the PDSs with the data collected on 2008 September 9
(MJD = 54718.3) and November 2 (MJD = 54772.2) in Figure \ref{power}. The
amplitude of QPO is calculated in the way of $rms_{\rm QPO} = (\pi \times LW
\times LN)^{1/2}$, where $LW$ is the full width half-maximum (FWHM) of the QPO,
and LN is the normalization of Lorentzian component for QPO. The parameters of
QPOs are listed in Table \ref{qpo_log}. Because of the large uncertainties, we
cannot describe the final six marginal QPO features in detail, and we
investigate only the first seven strong QPOs. The frequencies of these QPOs
were not only correlate with the intensity but also anti-correlated with the
hardness (Figure \ref{qpo}). In addition, the photon index $\Gamma$ and the
quality factor monotonically increased as the QPO frequency increased from 0.85
Hz to 5.31 Hz.

\section{Discussion}

It has been suggested that optical and NUV emissions in XRBs could stem from
different components \citep[e.g. ][]{rykoff07,weng15}. The modest correlation
between the U band and the X-ray fluxes (Figure \ref{lxluv}) implies that the U
band and the X-ray radiations have some connection via the X-ray reprocessing
or the tail of outer disk thermal emission. Meanwhile, the emission at longer
wavelengths does not show a significant correlation with the X-ray radiation,
pointing to some other contribution (e.g. the companion star or jet). In
particular, the U-band flux almost halved while the B-band flux enhanced by a
factor of $\sim 1.8$, when the source took an excursion in the HID (from MJD
54750 to 54755) (Figure \ref{lc}). At this time, the thermal and non-thermal
components have the comparable contribution to the (0.6--10 keV) X-ray
emission. Coincidentally, a weak QPO was detected on MJD 54753; however, the
low data statistic means that we cannot classify it. The positions in the HID
and HRD indicate that the source transited between the HIMS and the SIMS
(crossed the jet line), which could be associated to type-B QPOs and the launch
of relativistic jet \citep[see e.g. Figure 1 in ][]{kylafis15}. If the jet
produces part of the optical emissions, especially at long wavelengths, the
reverse trends shown at the shorter (U) and longer (B) bands can be explained.

Since early 2008 September, the X-ray spectra became softer with $\Gamma$
increasing from $\sim 1.5$ to $\sim 2.4$, indicating that \object{Swift
J1842.5-1124} started to transit from the LHS to HSS (via the HIMS). In the
meantime, the prominent QPO features were observed in the seven observations.
In addition to the position in the HID, the QPO frequencies, the high amplitude
of QPO, and the flat-top noise component displayed in the profiles of PDSs
(left panel of Figure \ref{power}) allow us to identify these QPOs as the
type-C QPOs.

Various approaches have been made to further our understanding of evolution of
low frequency QPOs in BH XRBs \citep[see e.g.][]{tagger99,titarchuk04,yan13}.
The truncated disk model considering the Lense-Thirring precession had been
proposed to explain the low frequency QPOs \citep[e.g. ][]{ingram09}, and now
has been developed to interpret the simultaneous observation of both high
frequency and low frequency QPOs \citep{motta14,fragile16}. In this model, the
QPO frequency is expected to increase as the truncation radius moves in.
Because of the low sensitivity of {\it RXTE}/PCA at below 2 keV, we cannot put
a constraint on the disk radius. Moreover, the irradiation effect at the LHS
would lead to an underestimate of disk radius, resulting in more uncertainties
\citep[e.g. ][]{gierlinski08}. Alternatively, when the source transited from
the LHS to the HSS, the centroid frequencies of the observed type-C QPO
monotonically increased with increasing intensity and decreasing hardness. The
level of fast variability became lower and the photon index $\Gamma$ also
softened, from 1.51 to 2.19 (Figure \ref{qpo}). All these correlations imply a
decreasing truncation radius and agree with the prediction of the
Lense-Thirring precession model.

\begin{acknowledgements}
We thank the referee for helpful comments that improved this work. This work is
supported by the National Natural Science Foundation of China under grants
11303022, 11133002, 11473023, 11543008, 11573023, 11173016 and 11433005, and by
the Special Research Fund for the Doctoral Program of Higher Education (grant
No. 20133207110006).
\end{acknowledgements}

%%%%%%%%%%%%%%%%%%%%%%%%%%%%%%%%%%%%%%%%%%%%%%%%%%%%%%%%%%%%%%%%%%%%%%%%%
%\bibliographystyle{aa}

{}

\end{document}